\documentstyle{article}

\font\tenrm=cmr10
\font\tenit=cmti10
\font\elevenbf=cmbx10 scaled\magstep 1
\font\elevenrm=cmr10 scaled\magstep 1

\renewenvironment{thebibliography}[1]
 { \tenrm
 \baselineskip=10pt
   \begin{list}{\arabic{enumi}.}
    {\usecounter{enumi} \setlength{\parsep}{0pt}
     \setlength{\itemsep}{3pt} \settowidth{\labelwidth}{#1.}
     \sloppy
    }}{\end{list}}

\begin{document}
\begin{center}{\elevenbf
FLAVOURED MULTISKYRMIONS. 
\\}
\vglue 0.5cm
{\tenrm V.B.Kopeliovich$^*$ and  W.J.Zakrzewski$^{**}$ \\}
{\tenit $^*$Institute for Nuclear Research of the Russian Academy of
Sciences, Moscow 117312, Russia\\
$^{**}$Department of Mathematical Sciences, University of Durham,
Durham DH1 3LE, UK\\}
\vglue 0.5cm
\end{center}
{\rightskip=3pc
\leftskip=3pc
\tenrm\baselineskip=10pt
\noindent
Static properties of multiskyrmions with baryon numbers up to 8 are
calculated, based on the recently given rational map ansaetze.
The spectra of baryonic systems with strangeness, charm and bottom are 
considered within a "rigid oscillator" version of the bound state soliton model.
It is suggested that the recently observed negatively 
charged nuclear fragment can be considered as a quantized strange
 multiskyrmion with $B=6$ or $7$.
In agreement with previous observation, baryonic systems with charm 
or bottom have more chance to
be bound by the strong interactions than strange baryonic systems. 
 \vglue 0.3cm}
 \vglue 0.5cm
\baselineskip=14pt
\elevenrm
1. The topological soliton models, and the Skyrme model among them \cite{1},
are attractive
because of their simplicity and the possibility that they may describe well
 various properties of low energy baryons. The models of this kind provide
also a very good framework within which to investigate the possibility
of the existence of nuclear matter fragments with unusual properties, 
such as flavour being different from
$u$ and $d$ quarks. 
In addition to being important by itself, this issue can have
consequences in astrophysics and cosmology. It is well known that
the relativistic many-body problems cannot be solved directly 
using the existing 
methods, and the chiral soliton approach may allow to overcome
some of these difficulties.

The description of skyrmions with large baryon numbers is complicated
because the explicit form of the fields was not known.
A recent remarkable observation \cite{2} that the fields
of the $SU(2)$ skyrmions can be 
approximated accurately by rational map ansaetze giving the values of 
masses close to their precise values, has simplified considerably their studies.
Similar ansaetze have also been recently presented for $SU(N)$ skyrmions
(which are not embeddings of $SU(2)$ fields)\cite{3}.

Here we use the $SU(2)$ rational map 
ansaetze as the starting points for the calculation of static
properties of bound states of skyrmions necessary for their quantization
in the $SU(3)$ collective coordinates  space.
The energy and baryon number densities of the $B=3$ configuration have 
tetrahedral symmetry, for $B=4$ - the octahedral
(cubic) one \cite{4}, for $B=5$ - $D_{2d}$-symmetry, for $B=6$ - $D_{4d}$, 
for $B=7$ - dodecahedral symmetry, and for $B=8$ - $D_{6d}$ - symmetry 
\cite{5,2}, etc.
The minimization, with the help of a 3-dimensional variational $SU(3)$ 
program \cite{6}, lowers the energies of these configurations by few 
hundreds of $Mev$
and shows that they are local minima in the $SU(3)$ configuration space.
The knowledge of the ``flavour" moment of inertia and the 
$\Sigma$-term allows us then to estimate the flavour excitation energies.
The mass splittings of the lowest states with different values
of strangeness, charm or bottom are calculated within the rigid oscillator
version of the bound state approach. The binding energies of baryonic systems
($BS$) with different flavours are also estimated.

2. Let us consider simple $SU(3)$ extentions of the Skyrme model \cite{1}: we 
start with $SU(2)$ skyrmions (with flavour 
corresponding to $(u,d)$ quarks) and extend them to various $SU(3)$ groups, 
$(u,d,s)$, $(u,d,c)$, or $(u,d,b)$.
We take the Lagrangian density of the Skyrme model, which in its well known 
form depends on parameters $F_{\pi}, \; F_D$ and $e$ and
can be written in the following way \cite{7}:
$${\cal L} =  \frac{F_\pi^2}{16}Tr l_{\mu} l^{\mu} + {1 \over 32e^2}
Tr [l_\mu,l_\nu]^2 +\frac{F_\pi^2m_\pi^2}{16} Tr(U+U^{\dagger}-2) + $$
$$ +\frac{F_D^2m_D^2-F_\pi^2m_\pi^2}{24}Tr(1-\sqrt{3}\lambda_8)(U+U^{\dagger}-2)+
 \frac{F_D^2-F_\pi^2}{48} Tr(1-\sqrt{3}\lambda_8)(Ul_\mu l^\mu +
l_\mu l^\mu U^{\dagger}).\eqno (1) $$
$U \in SU(3)$ is a unitary matrix incorporating chiral (meson) fields, and
$l_\mu=U^{\dagger} \partial _\mu U$. In this model $F_\pi$ is fixed 
at the physical value: $F_\pi$ = $186$ Mev. $M_D$ is the mass of $K, \, D$ or
$B$ meson. 

The flavour symmetry breaking $(FSB)$ in the Lagrangian $(1)$ is of
 the usual form, and was sufficient to describe the mass 
splittings of the octet and decuplets of baryons
\cite{7}. The Wess-Zumino term, not shown here, plays an important role in 
the quantization procedure, but it does not contribute to the
static masses of classical configurations \cite{8}.

We begin our calculations with $U \in SU(2)$, as was mentioned above.
The classical mass of $SU(2)$ solitons, in most general case, depends on $3$
profile functions: $f, \, \alpha$ and $\beta$.
The general parametrization of $U_0$ for an $SU(2)$ soliton we use here 
is given by $U_0 = c_f+s_f \vec{\tau}\vec{n}$ with $n_z=c_{\alpha}$, 
$n_x=s_{\alpha}
c_{\beta}$, $n_y=s_{\alpha}s_{\beta}$, $s_f=sinf$, $c_f=cosf$, etc.

The flavour moment of inertia enters directly in the procedure
of quantization \cite{9}-\cite{17}, and for arbitrary $SU(2)$ 
skyrmions is given by \cite{15,17}:
$$ \Theta_F = {1 \over 8} \int (1-c_f)\bigl[ F_D^2 + {1 \over e^2}
 \bigl( (\vec{\partial}f)^2 +s_f^2(\vec{\partial}\alpha)^2 +
s_f^2s_{\alpha}^2(\vec{\partial}\beta)^2\bigr) \bigr] d^3\vec{r}. \eqno (2) $$
It is simply connected with $\Theta_F^{(0)}$ for the flavour symmetric case:
$ \Theta_F=\Theta_F^{(0)}+(F_D^2/F_\pi^2-1) \Gamma/4,  $
$\Gamma$ is defined in $(3)$ below.
The isotopic moments of inertia are the components of the corresponding
tensor of inertia
\cite{9,10}, in our case this tensor of inertia is close to unit matrix
multiplied by $\Theta_T$.
The quantities $\Gamma$ (or $\Sigma$-term), which defines
the contribution of the 
mass term to the classical mass of solitons, and $\tilde{\Gamma}$ also 
are used in the quantization procedure:
$$ \Gamma = \frac{F_{\pi}^2}{2} \int (1-c_f) d^3\vec{r}, \qquad
  \tilde{\Gamma}={1 \over 4} \int c_f
 \bigl[ (\vec{\partial}f)^2 +s_f^2(\vec{\partial}\alpha)^2 +
s_f^2s_{\alpha}^2(\vec{\partial}\beta)^2 \bigr] d^3\vec{r}. \eqno (3)$$
The masses of solitons, moments of inertia, $\Gamma$ and $\tilde{\Gamma}$
 are presented in the Table below.
\vspace{1mm}
\begin{center}
\begin{tabular}{|l|l|l|l|l|l|l|l|l|l|l|l|}
\hline
 $B$  &$M_{cl}$& $\Theta_F^{(0)}$ & $\Theta_T$&$\Gamma$&$\tilde{\Gamma}$ &
$\omega_s$&$\omega_c$ &$\omega_b$&$\Delta \epsilon_s$&$\Delta \epsilon_c$
&$\Delta \epsilon_b$ \\
\hline
$1$&$1.702 $&$2.04$&$5.55$&$4.83$&$15.6$&$0.309$&$1.542$&$4.82$ 
&---&---&---  \\
\hline
$3$&$4.80 $&$6.34$&$14.4$&$14.0$&$27$&$0.289$&$1.504$&$4.75$  
&$-0.041$&$-0.01$&$0.03$ \\
\hline
$4$&$6.20 $&$8.27$&$16.8$&$18.0$&$31$&$0.283$&$1.493$&$4.74$ 
&$-0.020$&$0.019$&$0.06$ \\
\hline
$5$&$7.78$ &$10.8$&$23.5 $&$23.8$&$35$&$0.287$&$1.505$&$4.75$ 
&$-0.027 $&$0.006$&$0.05$ \\
\hline
$6$&$9.24 $&$13.1$&$25.4$&$29.0$&$38 $&$0.287$&$1.504$&$4.75$ 
&$-0.019 $&$0.017$&$0.05$ \\
\hline
$7$&$10.6 $&$14.7$&$28.7$&$32.3$&$44 $&$0.282$&$1.497$&$4.75$
&$-0.017$&$0.021$&$0.06$ \\
\hline
$8$&$12.2 $&$17.4$&$33.4$&$38.9$&$47$&$0.288$&$1.510 $&$4.77$ 
&$-0.018$&$0.014$&$0.02$ \\
\hline
\end{tabular}
\end{center}
\vspace{1mm}
{\baselineskip=8pt 
\tenrm
Characteristics of the bound states of skyrmions
with baryon numbers up to $B=8$. The classical mass of solitons $M_{cl}$ is 
in $Gev$, moments of inertia,
$\Gamma$ and $\tilde{\Gamma}$ - in $Gev^{-1}$, the excitation frequences 
for flavour $F$, $\omega_F$ in $Gev$. The parameters of the model
$F_{\pi}=186 \, Mev, \; e=4.12$. The accuracy of calculations is better
than $1\%$ for the masses and few $\%$ for other quantities. The $B=1$
quantities are shown for comparison. $\Delta \epsilon_{s,c,b}$, in $Gev$,
 are the changes of binding energies of lowest $BS$ with flavour 
$s,\,c$ or $b$, $|F|=1$, in comparison with usual $(u,d)$ nuclei (see Eq.(14)).\\}
\vspace{1mm}

3. To quantize the solitons in $SU(3)$ configuration space,  in the
spirit of the bound state approach to the description of strangeness
proposed in \cite{11,12} and used in \cite{13,14},
we consider the collective coordinates  motion of the 
meson fields incorporated into the matrix $U$:
$$ U(r,t) = R(t) U_0(O(t)\vec{r}) R^{\dagger}(t), \qquad
 R(t) = A(t) S(t), \eqno (4) $$
where $U_0$ is the $SU(2)$ soliton embedded into $SU(3)$ in the usual 
way (into the left upper corner), $A(t) \in SU(2)$ describes $SU(2)$ rotations,
$S(t) \in SU(3)$ describes 
rotations in the ``strange", ``charm" or ``bottom" directions, 
and $O(t)$ describes rigid rotations in real space.
$$ S(t) = exp (i{\cal D}(t)),\qquad
 {\cal D} (t)=\sum_{a=4,...7} D_a(t) \lambda_a, \eqno (5)$$
$\lambda_a$ are Gell-Mann matrices of the $(u,d,s)$, $(u,d,c)$ or $(u,d,b)$
$SU(3)$ groups. The $(u,d,c)$ and $(u,d,b)$ $SU(3)$ 
groups are quite analogous to
the $(u,d,s)$ one. For the $(u,d,c)$ group a simple redefiniton of 
hypercharge should be made. For the $(u,d,s)$ group,
 $D_4=(K^++ K^-)/\sqrt{2}$, $D_5=i(K^+-K^-)/\sqrt{2}$, etc.
And for the $(u,d,c)$ group $D_4=(D^0+\bar{D}^0)/\sqrt{2}$, etc.

The angular velocities of the isospin rotations are defined in the standard 
way: $ A^{\dagger} \dot{A} =-i\vec{\omega}\vec{\tau}/2.$
We shall not consider here the usual space rotations explicitly because the
corresponding moments of inertia for $BS$ are much greater than isospin
moments of inertia, and for lowest possible values of angular momentum $J$ the
corresponding quantum correction is either exactly 
zero (for even $B$), or small.

The field $D$ is small in magnitude, at least,
 of order $1/\sqrt{N_c}$, where $N_c$ is the number of colours in $QCD$.
Therefore, an expansion of the matrix $S$ in $D$ can be made safely. 
To the lowest order in field $D$ the Lagrangian of the model $(1)$
can be written as
$$L=-M_{cl,B}+4\Theta_{F,B} \dot{D}^{\dagger}\dot{D}-\biggl[\Gamma_B \biggl(
\frac{F_D^2}{F_{\pi}^2} m_D^2-m_{\pi}^2 \biggr)+ \tilde{\Gamma}_B
(F_D^2-F_\pi^2) \biggr] D^{\dagger}D -
 i{N_cB \over 2}(D^{\dagger}\dot{D}-\dot{D}^{\dagger}D). \eqno(7)$$
Here and below $D$ is the doublet $K^+, \, K^0$ ($D^0, \, 
D^-$, or $B^+,\,B^0$).
We have kept the standard notation for the moment of inertia of the
rotation into the ``flavour" direction $\Theta_F$ for $\Theta_s, \,
\Theta_c$ or $\Theta_b$ \cite{10,15} (the index $c$ denotes the 
charm quantum number, except in $N_c$). The contribution proportional to
$\tilde{\Gamma}_B$ is suppressed in comparison with the term $\sim \Gamma$
by the small factor $\sim (F_D^2-F_\pi^2)/m_D^2$, and is more important 
for strangeness.
The term proportional to $N_cB$ in $(7)$ arises from the Wess-Zumino term
in the action and is responsible for the difference of
the excitation energies of strangeness and antistrangeness 
(flavour and antiflavour in general case) \cite{13,14}.

Following the canonical quantization procedure the Hamiltonian of the 
system, including the terms of the order 
of $N_c^0$, takes the form \cite{11,12}:
$$H_B=M_{cl,B} + {1 \over 4\Theta_{F,B}} \Pi^{\dagger}\Pi + \biggl(\Gamma_B 
\bar{m}^2_D+\tilde{\Gamma}_B(F_D^2-F_\pi^2)+\frac{N_c^2B^2}{16\Theta_{F,B}} 
\biggr) D^{\dagger}D +i {N_cB \over 8\Theta_{F,B}}
(D^{\dagger} \Pi- \Pi^{\dagger} D). \eqno (8) $$
$\bar{m}_D^2 = (F_D^2/F_{\pi}^2) m_D^2-m_\pi^2$.
The momentum $\Pi$ is canonically conjugate to variable $D$.
Eq. $(8)$ describes an oscillator-type motion of the field $D$ 
in the background formed 
by the $(u,d)$ $SU(2)$ soliton. After the diagonalization which can be done
explicitly following \cite{13,14}, the normal-ordered Hamiltonian can be 
written as
$$H_B= M_{cl,B}+\omega_{F,B}a^{\dagger}a+\bar{\omega}_{F,B}b^{\dagger}b
 + O(1/N_c), \eqno (9) $$
with $a^\dagger$, $b^\dagger$ being the operators of creation of strangeness,
i.e., antikaons, and antistrangeness
(flavour and antiflavour) quantum number, $\omega_{F,B}$ and 
$\bar{\omega}_{F,B}$ being the 
frequences of flavour (antiflavour) excitations. $D$ and $\Pi$ are connected
with $a$ and $b$ in the following way \cite{13,14}:
$$ D^i= (b^i+a^{\dagger i})/\sqrt{N_cB\mu_{F,B}}, \qquad
\Pi^i = \sqrt{N_cB\mu_{F,B}}(b^i - a^{\dagger i})/(2i) \eqno (10) $$
with
$$ \mu_{F,B} =[ 1 + 16 (\bar{m}_D^2 \Gamma_B+(F_D^2-F_\pi^2)\tilde{\Gamma}_B)
 \Theta_{F,B}/ (N_cB)^2 ]^{1/2}. $$
For the lowest states the values of $D$ are small:
$ D \sim \bigl[16\Gamma_B\Theta_{F,B}\bar{m}_D^2 + N_c^2B^2 \bigr]^{-1/4}, $
and increase, with increasing flavour number $|F|$ like $(2|F|+1)^{1/2}$.
As was noted in \cite{14}, deviations of the field $D$ from the vacuum 
decrease with increasing mass $m_D$, as well as with increasing number of 
colours $N_c$, and the method works for any $m_D$ (and also for charm and 
bottom quantum numbers).

The excitation frequences $\omega$ and $\bar{\omega}$ are:
$$ \omega_{F,B} = N_cB(\mu_{F,B} -1)/(8\Theta_{F,B}), \qquad
 \bar{\omega}_{F,B} = N_cB(\mu_{F,B} +1)/(8\Theta_{F,B}).\eqno (11)$$
As was observed in \cite{15}, the difference
$\bar{\omega}_{F,B}-\omega_{F,B} = N_cB/(4\Theta_{F,B})$ coincides, to the 
leading order in $N_c$ with the expression obtained in the collective 
coordinates approach \cite{16}.

The $FSB$ in the flavour decay constants, i.e. the fact that $F_K/F_\pi 
\simeq 1.22$ and $F_D/F_\pi=1.7 \pm 0.2$ (we take $F_D/F_\pi=1.5$ and
$F_B/F_\pi=2$) leads to the increase of the flavour excitation 
frequences, in better agreement with data for charm and bottom \cite{18}. 
It also leads to some increase of the binding energies of $BS$ 
\cite{15}. 

The behaviour of static characteristics of multiskyrmions and flavour 
excitation frequences shown in the Table is similar to that obtained in
\cite{19} for toroidal configurations with $B=2,3,4$.
The flavour inertia
increases with $B$ almost proportionally to $B$. The frequences $\omega_F$
are smaller for $B \geq 3$ than for $B=1$.\\

4. The terms of the order of $N_c^{-1}$ in the Hamiltonian, which depend
 on the angular velocities of rotations in the isospin and the usual space and
which describe the zero-mode contributions are not crucial but important
for the numerical estimates of spectra of baryonic systems.

In the rigid oscillator model the states predicted do not correspond to
the definite $SU(3)$ or $SU(4)$ representations. How this can be remedied
 was shown in \cite{14}.  
For example, the state with $B=1$, $|F|=1$, $I=0$ should belong to the 
octet of $(u,d,s)$, or $(u,d,c)$, $SU(3)$ group, if $N_c=3$. 

Here we consider quantized states of $BS$ which belong to the lowest
possible $SU(3)$ irreps $(p,q)$, $p+2q=3B$: $p=0, \; q=3B/2$ for
even $B$, and $p=1, \; q=(3B-1)/2$ for odd $B$. For $B=3,\, 5$ and $7$ they 
are $35, \, 80$ and $143$-plets,
for $B=4, \, 6$ and $8$ - $28$,  $55$ and $91$-plets.
Since we are interested in the lowest energy states, we discuss here the
baryonic systems with the lowest allowed angular momentum,{\it ie} $J=0$,
for $B=
4, \; 6$ and $8$. For odd $B$ the quantization of $BS$ meets some difficulties,
but the correction to the energy of quantized states due to nonzero angular
momentum is small and decreases with increasing $B$ since the corresponding
moment of inertia increases proportionally to $\sim B^2$. Moreover, the
$J$-dependent correction to the energy cancels in the differences of energies
of flavoured and flavourless states which we discuss.

For the energy difference between the state with flavour $F$ belonging to the
$(p,q)$ irrep, and the ground state with $F=0$ and the same angular momentum
and $(p,q)$ we obtain:
$$ \Delta E_{B,F} = |F| \omega_{F,B} + \frac{\mu_{F,B}-1}{4\mu_{F,B}
\Theta_{F,B}}[I(I+1)-T_r(T_r+1)]
 + \frac{(\mu_{F,B}-1)(\mu_{F,B}-2)}{8\mu_{F,B}^2 \Theta_{F,B}} 
I_F(I_F+1), \eqno (12) $$
$T_r=p/2$ is the quantity analogous to the
``right" isospin $T_r$, in the collective coordinates approach \cite{9,10},
and $\vec{T_r}=\vec{I}_{bf}-\vec{I_F}$.
Clearly, the binding energy of multiskyrmions is cancelled in Eq. $(12)$.
For the states with maximal isospin $I=T_r+|F|/2$ 
the energy difference can be simplified to:
$$\Delta E_{B,F}=|F|\biggl[\omega_{F,B}+ T_r \frac{\mu_{F,B}-1}
{4\mu_{F,B}\Theta_{F,B}}
+\frac{(|F|+2)}{8\Theta_{F,B}} \frac{(\mu_{F,B}-1)^2}{\mu_{F,B}^2} \biggr].
\eqno (13) $$
This  difference depends on the flavour moment of inertia but not 
on $\Theta_{T}$. In the case of antiflavour excitations we have the same 
formulas, with the substitution $ \mu \to -\mu $. 
For even $B, \; T_r=0$, for odd $B$, $T_r=1/2$ for the lowest $SU(3)$ irreps.
It follows from $(12)$ and $(13)$ that when some nucleons are replaced by 
flavoured hyperons in $BS$ the binding energy of the system  changes by
$$\Delta \epsilon_{B,F}=|F|\biggl[\omega_{F,1}-\omega_{F,B} - \frac{3(\mu_
{F,1}-1)}{8\mu_{F,1}^2\Theta_{F,1}}- T_r \frac{\mu_{F,B}-1}{4\mu_{F,B}\Theta_{F,B}}
-\frac{(|F|+2)}{8\Theta_{F,B}} \frac{(\mu_{F,B}-1)^2}{\mu_{F,B}^2} \biggr]
\eqno (14) $$
For strangeness Eq. $(14)$ is negative indicating that stranglets should 
have binding energies smaller than those of nuclei, or can be unbound.
Since $\Theta_{F,B}$ increases with increasing $B$ and
$m_D$ this leads to the increase of binding with increasing $B$ and mass of
the "flavour", in agreement with \cite{15}. For charm and bottom Eq. $(14)$ 
is positive for $B \geq 3$, see the Table for the case $|F|=1$.

The nuclear fragments with sufficiently large values of strangeness (or bottom)
can be found in experiments as fragments with negative charge $Q$,
according to the
well known relation $Q=T_3+(B+S)/2$ (similarly for the bottom number).
One event of a long lived nuclear fragment with mass about
$7.4 Gev$ was reported in \cite{20}. Using the above formulas it is
not difficult to establish that this fragment can be the state with $B=-S=6$, 
or $B=7$ and strangeness $S=-3$. In view of some uncertainty of present
calculation - the rigid oscillator version of the model leads to overestimation
of flavour excitation energies - greater values of strangeness, by $1$ or $2$
units can be necessary to obtain the observed value of mass.

As in the $B=1$ case \cite{21} the absolute values of masses of
 multiskyrmions are controlled by the poorly
known loop corrections to the classic masses, or the Casimir energy. 
 And as was done for the $B=2$ states, \cite{16}, 
 the renormalization procedure is necessary to obtain physically reasonable
values of the masses. 
This generates an uncertainty of about several tens of $Mev$, 
as the binding energy of the deuteron is $30 \; Mev$ instead of the measured 
value $2.23 \; Mev$, so $\sim 30 \; Mev$ characterises the uncertainty of our 
approach \cite{16,17}. But this uncertainty is cancelled in the differences of
binding energies $\Delta \epsilon$ shown in the Table.\\

5. Using rational map ansaetze as starting configurations we have calculated
the static characteristics of bound skyrmions with baryon numbers up to $8$.
The excitation frequences for different flavours - strangeness, charm and
bottom - have been calculated using a rigid
oscillator version of the bound state approach of the chiral soliton
models. This variant of the model overestimates the mass splitting of
strange hyperons when $FSB$ in decay constant $F_K$ is included, but
works better for $c$ and $b$ flavours \cite{18}. 
Our previous conclusion that $BS$ with charm and bottom have more chances
to be bound respectively to strong decay than strange $BS$ \cite{15} is 
reinforced by the present investigation. This conclusion takes place also
in $FS$ case, $F_D=F_\pi$.

Consideration of the $BS$ with ``mixed" flavours is  possible in principle, 
but would be technically more involved. Our results agree qualitatively 
with the results of \cite{22} where the strangeness excitation frequences 
had been calculated within the bound state approach. The difference is, 
however, in the behaviour of excitation frequences: we have found that they 
decrease when the baryon number increases
from $B=1$ thus increasing the binding energy of corresponding $BS$.

The charmed baryonic systems with $B=3,\,4$ were considered in \cite{23}
within a potential approach. The $B=3$ systems were found to be
very near the threshold and the $B=4$ system was found to be stable 
with respect to the strong decay, with a binding energy of $\sim 10 \, Mev$.
Further experimental searches for the baryonic systems with flavour 
different from $u$ and $d$ could shed more light on the dynamics of 
heavy flavours in baryonic systems.

This work has been supported by the UK PPARC grant: PPA/V/S/1999/00004.\\
VBK is indebted to the Institute for Nuclear Theory at
the University of Washington where the work was initiated 
for its hospitality and DOE for support. 

\end{document}